\documentclass[aps,pra,reprint,amsmath,amssymb,superscriptaddress,nobibnotes,showpacs,showkeys]{revtex4-1}

\usepackage{dcolumn}
\usepackage{bm}
\usepackage{amsmath}
\usepackage{amssymb,amsthm}
\usepackage{bbm}
\usepackage{epsfig,color}
\usepackage[sans]{dsfont}
\usepackage{comment}
\usepackage{hyperref}
\usepackage{units}
\usepackage{color}
\usepackage{ifthen}
\usepackage{graphicx}
\graphicspath{ {images/} }

\definecolor{nblue}{rgb}{0.2,0.2,0.7}
\definecolor{ngreen}{rgb}{0.2,0.6,0.2}
\definecolor{nred}{rgb}{0.7,0.2,0.2}
\definecolor{nblack}{rgb}{0,0,0}

\newcommand{\ket}[1]{|#1\rangle}

\newcommand{\tr}{\text{tr}}

\renewcommand{\H}{\mathcal{H}}

\def\B{\mathcal{B}}

\newcommand{\N}{\mathcal{N}}
\newcommand{\x}{\textrm{x}}
\newcommand{\D}{\mathcal{D}}
\newcommand{\g}{\mathcal{g}}

  \theoremstyle{definition}
  
  \theoremstyle{plain}
  
\theoremstyle{plain}
  
\theoremstyle{plain}

  \theoremstyle{plain}
  
  \theoremstyle{plain}

  \providecommand{\conjecturename}{Conjecture}
  \providecommand{\definitionname}{Definition}
  \providecommand{\lemmaname}{Lemma}
\providecommand{\corollaryname}{Corollary}
\providecommand{\theoremname}{Theorem}
\providecommand{\propositionname}{Proposition}

\def\x{\mathrm{x}}
\def\y{\mathrm{y}}

\def\i{\mathrm{id}}

\def\g{\mathrm{guess}}

\def\N{\mathcal{N}}

\def\H{\mathcal{H}}

\def\M{\mathrm{M}}

\def\D{\mathcal{D}}

\def\tr{\mbox{tr}}

\def\bea{\begin{eqnarray}}
\def\eea{\end{eqnarray}}

\begin{document}

\title {  Channel Coding of a Quantum Measurement }

\author{Spiros Kechrimparis}
\affiliation{School of Electrical Engineering, Korea Advanced Institute of Science and Technology (KAIST), 291 Daehak-ro Yuseong-gu, Daejeon 34141 Republic of Korea,}
\author{ Chahan M. Kropf}
\affiliation{ Istituto Nazionale di Fisica Nucleare, Sezione di Pavia, via Bassi 6, I-27100 Pavia, Italy, }
\affiliation{Dipartimento di Matematica e Fisica and Interdisciplinary Laboratories for Advanced Materials Physics, Universit\`{a} Cattolica del Sacro Cuore, via Musei 41, I-25121 Brescia, Italy,}
\author{ Filip Wudarski}
\affiliation{Quantum Artificial Intelligence Lab. (QuAIL), Exploration Technology Directorate, NASA Ames Research Center, Moffett Field, CA 94035, USA,}
\affiliation{Research Institute for Advanced Computer Science, 615 National Ave., Suite 220, Mountain View, CA 94043, USA,}
\affiliation{Institute of Physics, Faculty ofPhysics, Astronomy and Informatics, Nicolaus Copernicus University, Grudziadzka 5/7, 87?100 Torun, Poland.}
\author{ Joonwoo Bae }
\affiliation{School of Electrical Engineering, Korea Advanced Institute of Science and Technology (KAIST), 291 Daehak-ro Yuseong-gu, Daejeon 34141 Republic of Korea,}

\begin{abstract}
In this work, we consider the preservation of a measurement for quantum systems interacting with an environment. Namely, a method of preserving an optimal measurement over a channel is devised, what we call channel coding of a quantum measurement in that operations are applied before and after a channel in order to protect a measurement. A protocol that preserves a quantum measurement over an arbitrary channel is shown only with local operations and classical communication without the use of a larger Hilbert space. Therefore, the protocol is readily feasible with present day's technologies. Channel coding of qubit measurements is presented, and it is shown that a measurement can be preserved for an arbitrary channel for both i) pairs of qubit states and ii) ensembles of equally probable states. The protocol of preserving a quantum measurement is demonstrated with IBM quantum computers. 
\end{abstract}

\maketitle

Quantum systems are generally fragile in that they often interact with an environment \cite{ref:breuer, ref:alicki}. One of the consequences is that a designed quantum information task becomes noisy. For instance, when quantum states are stored in a memory, they may interact with an environment so that the resulting noisy states are finally read out by a measurement. In a communication scenario, while quantum states are transmitted, interactions with an environment take place:  a state sent by a party suffers from interventions of an environment, and then is measured by a receiver. 

The aforementioned scenario shares similarities with noisy channels in information theory \cite{ref:hamming}. After messages are encoded, sequences are transmitted and then corrupted during the transmission by a noisy channel. In information theory, the problem of unwanted interactions with an environment is resolved by channel coding, in which messages are prepared in longer sequences with additional bits in order to contain some redundancy on purpose such that the redundant bits are used to detect and correct errors that have appeared during the transmission. 

In quantum information theory, in a similar vein, quantum systems can be protected from unwanted interactions with an environment by exploiting more resources in the state preparation. As quantum states are described by linear, non-negative, and unit-trace operators in a Hilbert space, quantum states can be prepared in a subspace, also called as a code space, of a larger Hilbert space such that the complementary subspace is used to detect and correct errors that occurred in the code space. Consequently, the states prepared in a code state can be protected while systems suffer from interactions with an environment. The schemes have been referred to as quantum error correction \cite{shor1995,calderbank1996} or noiseless subsystems \cite{kribs2005}.

So far, in both cases of classical and quantum scenarios that deal with unwanted interactions with an environment, the goal is to preserve systems prepared in sequences or states, respectively. In quantum theory, we recall the significance of an optimal measurement to read out which state a system is prepared in. The role of a quantum measurement is illustrated as follows. Suppose that there are two sets of states, $S_Z = \{|0\rangle, |1\rangle \}$ and $S_X = \{|+ \rangle, |-\rangle \}$, where $|\pm \rangle = (|0\rangle \pm |1\rangle)/\sqrt{2}$. Although both contain a pair of orthogonal states, the perfect distinguishability can be achieved only when a measurement is performed in a correct basis: the $Z$ basis for the set $S_Z$ and the $X$ basis for the set $S_X$.  Measurements in the $Z$ basis for the states $S_X$, or $X$ to states $S_{Z}$, give no information to discriminate between the pair of orthogonal states. 

It is clear that if states are protected by a channel, so is an optimal measurement prepared for the states. However, the preservation of a measurement does not necessarily imply that quantum states should be protected completely. For instance, suppose that two states $\{|0\rangle, |1\rangle \}$ are sent through a channel 
\bea
|0\rangle &\mapsto& (1-p) |0\rangle \langle 0| + p |1\rangle \langle 1|, \nonumber \\
|1\rangle &\mapsto& (1-p) |1\rangle \langle 1| + p |0 \rangle \langle 0| \label{eq:2} 
\eea 
for $p \in [0,1]$. An optimal measurement for state discrimination remains the same as a measurement in the $Z$ basis for both before and after a channel use. The measurement in the $Z$ basis is optimal for both the initial ensemble and its resulting ensemble. That is, a measurement can be preserved over a channel that contains interactions with an environment. This shows that the preservation of a measurement is not equivalent to the preservation of states. It is then natural to ask if a quantum measurement can be preserved over a channel in general although states may become noisier by the channel. 

The preservation of an optimal measurement is also useful in a practical point view: one may desire to exploit a measurement setting, once prepared, repeatedly in a different environment. This happens, for instance, in quantum algorithms: a measurement in the computational basis is supposed to find a solution from a resulting state after all. If states cannot be fully preserved due to noise from an environment, a next best option could be to find an optimal measurement for a noisy resulting state, in order to maximize the probability of obtaining a solution. If the optimality of a measurement is preserved, a measurement does not have to be revised but remains optimal ever. Similarly, in quantum communication where states are sent through a noisy channel, an optimal measurement can find which state has been sent through a channel, although states may not be protected completely. The scheme of preserving an optimal measurement over a channel can be referred to as {\it channel coding of a quantum measurement}. We recall that methods of preserving states have been called channel coding of states \cite{ref:shannon}. 

The advantages of preserving an optimal measurement over a quantum channel are twofold. Firstly, the verification of resulting states can be circumvented. If a measurement once prepared would be optimal ever after a channel use, the step of identifying resulting states can be bypassed: even if the states are unknown, an optimal measurement for the states has been immediately there. In this way, quantum tomography that is highly demanding in practice can be circumvented. This can also be interpreted that measurement devices do not have to be realigned under unknown and unwanted interactions with an environment. For instance, measurement devices prepared in a laboratory can be re-used in some other applications such as platforms of satellite or free-space quantum communication, in which it is in fact difficult to characterize the environment. Next, as it is shown above, the preservation of an optimal measurement implements a cost effective optimal scheme of extracting information from quantum states. 

\begin{figure}[!t]
\centering
\includegraphics[width=3.4in]{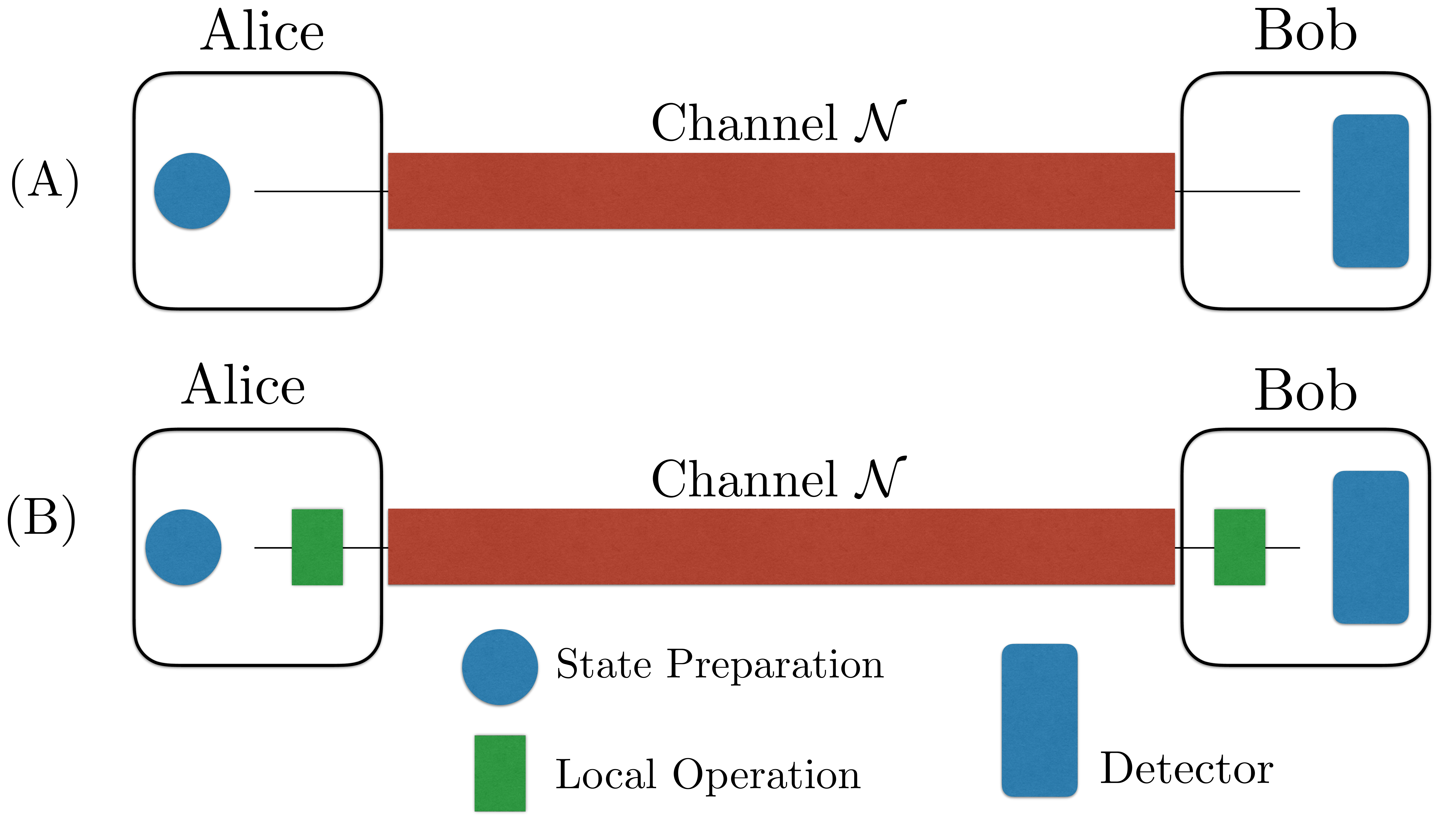}
\caption{ Quantum information processing consists of preparation, channel evolution, and measurements of quantum states. Channel coding can protect states or a measurement against interactions with an environment during the transmission. By channel coding of a measurement, detectors prepared in a noiseless scenario can be used repeatedly when states are sent through an arbitrary and unknown channel $\N$.}
	\label{fig:scheme}
\end{figure}

In this work, we show a framework of channel coding of a measurement by local operations and classical communication (LOCC), without resort to a larger Hilbert space. By chracterizing channels that preserve a measurement for an ensemble, channel coding of a measurement is formulated as a supermap from a channel to a measurement-preserving one, where a supermap can be implemented by an LOCC protocol. We present channel twirling, which is implemented by a unitary $2$-design, as channel coding of a measurement for ensembles of equally probable states in general. We then consider channel coding of a qubit measurement for i) any pair of states and ii) ensembles of equally probable states. Proof-of-principle demonstrations are shown with IBM quantum computers.

\section{ Preliminaries }

Let us begin with notations and terminologies to be used throughout. The building blocks of quantum information processing, states, channels, and measurements are summarized, see e.g., \cite{ref:chuang, ref:wilde}. Discrimination of quantum states under a channel is introduced. 

\subsection{States, channels, and measurements }

Let $\B(\H)$ denote a set of bounded linear operators in a Hilbert space $\H$. A set of quantum states on a Hilbert space $\H$ is denoted by $S(\H)$, i.e., 
\bea 
S(\H) = \{ \rho \in \B(\H) ~:~\tr[\rho] =1,~\rho\geq 0 \}. \nonumber
\eea
A quantum channel that describes transformations of quantum states, denoted by,
\bea
\N : \rho \mapsto \N[\rho] \nonumber
\eea
is characterized by a completely positive and trace preserving map for quantum states, i.e., $\i \otimes \N\geq 0$, and $\tr[\N[\rho]] =1$, $\forall \rho \in S(\H)$ where $\i$ is an identity map. 

A quantum measurement that shows the transition from quantum states to measurement outcomes contains positive operator values measure (POVM) elements, denoted as,
\bea 
\{ M_k\}_{k=1}^n, ~\mathrm{such~that}~ M_{k} \geq 0 ~\forall k, ~ \sum_{k} M_k = {\bf I}.\nonumber
\eea 
Note that each POVM element gives the description of a detector. The measurement postulate states that when a system is prepared in a state $\rho$, the probability of having a detection event on a POVM element $M_k$ is given by $\mathrm{Prob}[ k] = \tr[\rho M_k]$. 

\subsection{Optimal quantum state discrimination over a channel}

Discrimination of quantum states is a fundamental task that a measurement finds which state has been prepared \cite{holevo1974,helstrom1976,yuen1975}. Let $S^{(\i)}$ denote a set of quantum states in a noiseless environment. We also introduce its noisy ensemble, denoted by $S^{(\N)}$, i.e., the set of resulting states after a channel $\N$. That is, we write the ensembles as follows,  
\bea
S^{(\i)} & = & \{ q_{\x}, \rho_{\x}  \}_{\x=1}^n ~\mathrm{and} \nonumber \\
S^{(\N) } & = &\{ q_{\x}, \N[ \rho_{\x} ] \}_{\x=1}^n, \label{eq:ens}
\eea
where $q_{\x}$ denotes the {\it a priori} probability that a state $\rho_{\x}$ is prepared. In addition, we write an ensemble of equally probable states as follows,
\bea
S_{0}^{(\i)} = \{ 1/n, \rho_{\x} \}_{\x=1}^n. \label{eq:eens}
\eea
Throughout, $S^{(\i)}$ means an ensemble of states prepared in a noiseless environment, and $S^{(\N)}$ the resulting ensemble by a channel $\N$.

The problem of optimal state discrimination finds the maximal probability of making a correct guess about which state it is, called the guessing probability, as well as an optimal measurement that attains the guessing probability \cite{bae2013,chefles2000,bae2013b,barnett2009,bae2015,bergou2004,bergou2007,bergou2010}. For the generality, suppose that states are sent through a channel $\N$. The guessing probability for the ensemble $S^{ ( \N) }$ can be written as, 
\bea
p_{\g}^{(\N)} = \max_{\M} \sum_{\x } q_{\x} \tr[ \N[ \rho_{\x}] M_{\x}] \label{eq:guessp}
\eea
where $\M$ denotes a set of POVM elements. 

The standard problem of optimal state discrimination corresponds to the case $S^{(\i)}$, for which the guessing probability is denoted by $p_{\g}^{(\i)}$. It is clear that $ p_{\g}^{(\i)} \geq p_{\g}^{(\N)}$ for a channel $\N$ \cite{kechrimparis2018}, i.e., distinguishability does not increase under a quantum channel. Note also that optimal POVM elements are not always non-zero, i.e., no-measurement is sometimes optimal \cite{ref:hunter}. 

\section{ Preservation of quantum resources} 

We summarize known results on the preservation of quantum states and a measurement. 

\subsection{The preservation of states}

Suppose that one of the states in $S^{(\i)}$ is sent through a channel $\N$. Quantum states can be preserved if there exists a recovery map $\mathcal{R}$ such that $ \mathcal{R \circ N } \approx \i$. If a recovery map does not exists, a larger Hilbert space can be exploited such that states are then encoded in its subspace, called a code space, for which there exists a recovery map. The complementary subspace can be used to detect and correct errors occurred in the code space. Or, states can be encoded in a subspace where the states are not affected by interactions with an environment \cite{shor1995,calderbank1996}.

The aforementioned methods of preserving quantum states can be generally characterized by the information preserving structure \cite{kohout2008}. Namely, quantum states $\{ \rho_{\x} \}_{\x=1}^n$ are preserved by a channel $\N$ if and only if the following conditions are satisfied, for all $q\in(0,1)$, 
\bea
\forall \x, \y,~ \| q \rho_\x - (1-q) \rho_{\y}   \|_1 = \| q \N [\rho_\x ] - (1-q) \N [\rho_{\y} ] \|_1  ~~~~\label{eq:ips}
\eea
where $\| \cdot \|_1$ denotes the $L_1$ norm. The information preserving structure can be used to construct the code states which remain perfectly distinguishable after a channel use.

\subsection{ The preservation of a measurement}

The channels that preserve an optimal measurement for an ensemble can be defined as follows. \\

{\bf Definition.} A channel $\N$ is called optimal measurement preserving (OMP) for an ensemble $S^{(\i)}$ if the resulting ensemble $S^{(\N)}$ and the ensemble $S^{(\i)}$ share the same measurement for optimal state discrimination. \\

In Ref. \cite{kechrimparis2018}, it has been shown that a channel $\N$ is OMP for an ensemble $S^{(\i)}$ if the following is satisfied, 
\bea
\forall \x,\y,~~  ( q_{\x} \rho_{\x}  - q_{\y} \rho_{\y} )=  \kappa^{-1}~ ( q_{\x}\N [\rho_\x ] -q_{\y} \N [\rho_{\y} ] )   ~\label{eq:popt}
\eea
for some $\kappa \in (0,1]$. The condition shares some similarities with the information preserving structure in Eq. (\ref{eq:ips}). 


Technically, taking the $L_1$ norm in the OMP condition in Eq. (\ref{eq:popt}) for $\kappa=1$, one can obtain the condition of preserving states in Eq. (\ref{eq:ips}) with a specific set of {\it a priori} probabilities $\{q_{\x} \}_{\x=1}^n$. If states can be preserved such that the condition in Eq. (\ref{eq:ips}) holds true for arbitrary probabilities $\{q_{\x} \}_{\x=1}^n$, there exists a recovery operation that finds the states before a channel use \cite{kohout2008}: $\exists \mathcal{R}$ such that $\mathcal{R} \circ\N [\rho_{\y} ] = \rho_{\y}$, for $\rho_{\y} \in \mathrm{span}\{\rho_{\x}  \}_{\x=1}^n$. This shows that, when a set of states can be preserved by a channel, i.e., Eq. (\ref{eq:ips}) is fulfilled, it follows that an optimal measurement can also be preserved, i.e., the OMP condition is satisfied, by applying a recovery operation. However, the OMP condition with $\kappa <1$ cannot be reduced to the information preserving structure for the preservation of states. This means that whereas the preservation of quantum states is in failure, an optimal measurement can be preserved.

As an example, let us consider ensembles of equally probable states $S_{0}^{(\i)}$ in Eq. (\ref{eq:eens}), for which channels in the following are OMP. \\

{\bf Remark.} A channel $\N_{\sigma} : \rho \mapsto (1-\eta) \rho + \eta \sigma$ for a fixed state $\sigma$ is OMP for any ensemble $S_{0}^{(\i)}$. \\

It is clear that channels $\N_{\sigma}$ do not preserve states. One can easily find that a channel $\N_{\sigma}$ satisfies the OMP condition in Eq. (\ref{eq:popt}). A particular choice of the state $\sigma$ can be a mixed state ${\bf I}/d$ where $d$ denotes the dimension of a Hilbert space. This introduces a depolarization channel as follows, 
\bea
D_{\eta} [\rho] = (1-\eta) \rho + \eta  {\bf I}/d.  \label{eq:dep}
\eea
A depolarization channel is then OMP for any ensemble $S_{0}^{(\i)}$.

\section{Channel coding of a measurement}
\label{sec:chm}

We now formulate channel coding of a quantum measurement by devising a transformation of a channel to an OMP one by an LOCC protocol. As channels are transformed into each other, such a transformation corresponds to a supermap \cite{chiribella2008} from a channel $\N$ to an OMP one. \\

{\bf Definition. (Channel coding of a measurement)} For a quantum channel $\N : S(\H) \rightarrow S(\H)$ and an ensemble $S$ of interest to be sent through the channel, let $\mathcal{C}^{(\N,S)}$ denote a supermap that works as $\N \mapsto \mathcal{C}^{(\N,S)} [ \N ]$. Then, a supermap $\mathcal{C}^{(\N,S)}$ is called channel coding of a measurement if the channel $\mathcal{C}^{(\N,S)} [ \N]$ is OMP for the ensemble $S$. \\

This shows that for the construction of channel coding of a measurement, one first needs to identify an OMP channel, to which a channel is to be transformed. The next step is then to find an LOCC protocol, i.e., a supermap, that transforms a channel to the OMP channel. 

We here take a depolarization channel in Eq.\@ (\ref{eq:dep}) as an OMP channel of interest. In this way, one can take the advantage of exploiting the well-known result that any channel can be transformed to a depolarization map by the protocol of twirling a channel, which also applies LOCC only \cite{horodecki1999}.

 Let $\mathcal{T}$ denote a twirling operation for a channel $\N$: 
\bea 
\label{eq:twirling}
\mathcal{T} \N [\rho] = \int d\mu (U) U^\dagger \N [ U \rho U^\dagger] U \,,
\eea
where the average is performed over the Haar measure (uniform measure in the space of unitary operators). The consequence is that the resulting map corresponds to a depolarization, 
\bea
\label{eq:twirlingDepol}
\mathcal{T} \N [\rho] = D_{\eta_{\N}} [\rho ] = (1-\eta_{\N} )\rho + \eta_{\N} \frac{ {\bf I}}{d}, \label{eq:depn}
\eea
where $\eta_{\N}$ is determined by a channel $\N$. Note also that $D_{\eta_{\N}}$ is a quantum channel for $1-\eta_{\N} \in [ - 1 / ( d^2-1) ,1]$.

In practice, twirling a channel can be realized by the use of a so-called \emph{unitary 2-design}.  A unitary 2-design is a set of unitary transformations in a $d$-dimensional Hilbert space, denoted by a set $ W= \{U_k\}_{k=1\ldots N}$, such that the following is satisfied \cite{gross2007}. For any quantum channel $\N$, it holds that 
\bea
		\frac{1}{N}\sum_{i=1}^N U_i^\dagger \N [ U_i \rho U_i^\dagger] U_i = \int_{U(d)} dU \, U^\dagger \N [ U \rho U^\dagger] U \,.\nonumber
\eea
Then, twirling a channel can be realized by random applications of a unitary $2$-design. 

Therefore, by applying a unitary $2$-design before and after a channel, channel coding of a measurement for an ensemble of equally probable states $S_{0}^{(\i)}$ can be implemented as follows, see also Fig. \ref{fig:scheme}. \\ 

{\bf A protocol of channel coding of a measurement}
\begin{enumerate}
\item For a state $\rho \in S_{0}^{(\i)}$, an element $U_j \in  W $ is randomly chosen from a unitary $2$-design $W$ and applied to the state before a channel $\N$. 
\item The sender and the receiver communicate their selection of unitaries. 
\item After the channel transmission, the receiver applies its inverse $U_{j}^{\dagger}$ to a resulting state $\N [ U_j \rho U_{j}^{\dagger }]$. 
\item By randomizing the resulting states, a measurement to find a state in an ensemble $S_{0}^{(\i)}$ is also optimal for its resulting ensemble $S_{0}^{(\N)}$ for any quantum channel $\N$.\\
\end{enumerate}

We remark that with the protocol, a measurement once prepared for an ensemble $S_{0}^{(\i)}$ remains ever optimal no matter what interactions a system suffers from an environment. Distinguishability does not increase under channels. Let $p_{\g}^{(\i)}$ denote the guessing probability for an ensemble $S_{0}^{(\i)}$,  $p_{\g}^{(\N)}$ for the ensemble $S_{0}^{(\N)}$, and  $p_{\g}^{(\mathcal{TN})}$ for the ensemble $S_{0}^{( \mathcal{TN})}$. It is clear that 
\bea
p_{\g}^{(\i)} \geq p_{\g}^{(\N)} ~\mathrm{and}~ p_{\g}^{(\i)} \geq p_{\g}^{( \mathcal{TN})}\nonumber
\eea 
since distinguishability does not increase under a channel. However, distinguishability can be improved by channel coding of a measurement \cite{kechrimparis2018}. Namely,
\bea
\exists \N,~~\mathrm{such~that~} p_{\g}^{(\N)} < p_{\g}^{(\mathcal{TN})}.\nonumber
\eea
Note also that resources to realize the protocol of channel coding of a measurement contain LOCC only: local unitaries are only applied before and after a channel use, which are feasible with current technologies.

\section{ Channel coding of a qubit measurement} 

From the framework of channel coding of a measurement in the previous section, we here consider channel coding of a qubit measurement. As it is shown, a unitary $2$-design is an essential tool to implement channel coding of a quantum measurement by transforming a channel to a depolarization map.  

\subsubsection{Unitary $2$-design}

For qubit cases, a  subgroup of the tetrahedral group of rotations \cite{bengtsson2006} that forms a unitary $2$-design with a conjectured minimal \cite{gross2007} cardinality of $12$ is chosen. Then, these elements can be found explicitly as follows:  
\bea
&& W = \{{\bf I},-i X, -i Y, -i Z , U_5,\cdots, U_{12} \} \label{eq:u2} 
\eea
where $X$, $Y$, and $Z$ are the Pauli matrices, and
\bea
&& U_5 = \frac{1}{2}\begin{pmatrix} 1-i & -1-i \\ 1-i & 1+i \end{pmatrix},  U_6 = \frac{1}{2}\begin{pmatrix} 1+i & 1-i \\ -1-i & 1-i \end{pmatrix},  \nonumber \\
&& U_7 = \frac{1}{2}\begin{pmatrix} -1-i & -1-i \\ 1-i & -1+i \end{pmatrix},  U_{8} = \frac{1}{2}\begin{pmatrix} -1+i & 1-i \\ -1-i & -1-i \end{pmatrix},  \nonumber \\
&& U_{9} = \frac{1}{2}\begin{pmatrix} -1+i & -1+i \\ 1+i & -1-i \end{pmatrix},  U_{10} = \frac{1}{2}\begin{pmatrix} -1-i & 1+i \\ -1+i & -1+i \end{pmatrix} \nonumber\\ 
&& U_{11} =  \frac{1}{2}\begin{pmatrix} 1+i & -1+i \\ 1+i & 1-i \end{pmatrix}, U_{12} = \frac{1}{2}\begin{pmatrix} 1-i & 1+i \\ -1+i & 1+i \end{pmatrix}.\nonumber
\eea
With the latter unitary $2$-design, the protocol in Sec. \ref{sec:chm} can realize channel coding of a measurement for ensembles of equal {\it a priori} probabilities $S_{0}^{(\i)}$, as it transforms a channel to a depolarization map. An alternative and widely used unitary $2$-design is the Clifford group that contains $24$ elements up to phase factors.

\subsubsection{Validity of channel coding of a measurement} Channel coding of a measurement is closely related to the validity of a measurement for optimal state discrimination. This is because no-measurement, i.e., a single POVM element is given by ${\bf I}$, is sometimes optimal, which we call a measurement a trivial measurement. In this case, an optimal discrimination is simply to make a guess according to {\it a priori} probabilities. Channel coding of a measurement is not valid if a channel $\mathcal{TN}$ leads to an ensemble $S^{(\mathcal{TN})}$ for which a measurement is trivial. 

Furthermore, a caveat is also that from Eq. (\ref{eq:depn}), the preservation of a measurement does not work if a resulting depolarization map has $1-\eta_{\N}<0$ since states of an ensemble $S^{(\i)}$ are less probable than their complement ones after a channel use. Note that for qubits, we have $1-\eta_{\N} \in [-1/3,1]$. Then, for $1-\eta_{\N} \in [-1/3,0]$, it is still possible to systematically re-construct an optimal measurement for a resulting ensemble without further efforts or resources such as tomography of a channel or states. We call the update of a measurement as a pre-protocol for quantum communication.

In what follows, we first describe a pre-protocol as a method of updating an optimal measurement. Then, we show that the protocol in Sec. \ref{sec:chm} works for an arbitrary pair of qubit states even if {\it a priori} probabilities are unequal. That is, twirling channel can preserve an optimal measurement in two-state discrimination in general. The sufficient condition in Eq. (\ref{eq:popt}) is also necessary to preserve a measurement for a pair of states. However, for more than two states, the condition is only sufficient as there are ensembles of three states with unequal {\it a priori} probabilities, such that the protocol in Sec. \ref{sec:chm} does not preserve a measurement. A counter-example is also provided. 

\subsection{Pre-protocol for communication}

\subsubsection{No-measurement is sometimes optimal}

Let us begin with discrimination of two states. Let $S^{(\i)}=\{q_\x,\rho_\x\}_{\x=1,2}$ denote a pair of qubit states. A measurement for optimal discrimination can be found by finding the spectrum of the operator, $q_1\rho_1 - q_2 \rho_2$ : two projectors into positive and negative eigenvalues constitute an optimal measurement. As it is mentioned, this construction is valid only when a measurement is non-trivial. In fact, an optimal measurement is non-trivial only when the following condition is satisfied \cite{weir2017}
\bea
| q_1-q_2 | < \| q_1 \rho_1 -q_2 \rho_2 \|_1.  \label{eq:con1}
\eea
Otherwise, an optimal measurement is trivial, i.e., one of the POVM elements is zero and the other is the identity. The optimal discrimination is achieved by guessing the state that has a higher {\it a priori} probability, i.e., the guessing probability is therefore given by $p_{\g} = \max \{ q_1, q_2 \}$. 

Together with channel coding of a measurement, the protocol in Sec. \ref{sec:chm} performs twirling a channel $\N$ and the resulting channel $D_{\eta_{\N}}$ is obtained. Applying the condition in Eq. (\ref{eq:con1}), it follows that a measurement is trivial if
\bea
| q_1 - q_2 | >  |1-\eta_{\N} |  \| q_1 \rho_1 -q_2 \rho_2 \|_1 . \nonumber
\eea
In this case, the optimal discrimination is to guess a state according to {\it a priori} probabilities. Otherwise, the protocol in Sec. \ref{sec:chm} can preserve a measurement.

Recently \cite{weir2017}, it has been shown that for an ensemble $S^{(\i)}$ of multiple states, a measurement is trivial if there exists a state $\rho_j$ such that,
\bea
| q_j-q_k | > \| q_j \rho_j -q_k \rho_k \|_1,~~\forall k. \nonumber
\eea
Thus it follows that an optimal measurement after a channel $\N$ is trivial if there is a state $\rho_j$ such that
\bea
| q_j-q_k | > (1-\eta_{\N}) \| q_j \rho_j -q_k \rho_k \|_1,~~\forall k. \nonumber
\eea
The preservation of a measurement is valid when a measurement is non-trivial in both before and after a channel use. 

\subsubsection{Updating a measurement} Suppose that after twirling a channel, a depolarization map $D_{\eta_{\N}}$ has the noise fraction $1-\eta_{\N} <0$, see Eq. (\ref{eq:depn}). This may be compared to its classical countepart, a binary symmetric channel \cite{nielsen2000} with $1-p <1/2$ where $p$ denotes the probability of flip: then $\x$ is mapped to $\x \oplus 1$ more frequently than $\x$, where $\oplus$ is the bitwise addition. In the classical case, by re-labeling $\x$ to $\x+1$ before or after a binary symmetric channel, the probability of flip can be suppressed to be smaller than $1/2$. 

Similarly, if $1-\eta_{\N} <0$, states in an ensemble $S^{(\i)}$ are less probable than their orthogonal complement : a measurement is not preserved. The update of an optimal measurement is shown as follows. \\

{\bf Proposition.} For an ensemble $S_{0}^{(\i)}$, let $\{M_{\x} \}_{\x=1}^n $ denote an optimal measurement. POVM elements can be written in terms of the Bloch vectors, $M_{\x} = w_{\x} s (\vec{m}_{\x}) $ such that $w_{\x} \in [0,1]$ and $s (\vec{m}_{\x}) = ({\bf I} + \vec{m}_{\x} \cdot \vec{\sigma})/2$ where $\vec{\sigma} = (X,Y,Z)$ with Pauli matrices. For a channel $\N$ having $1-\eta_{\N} <0 $ after channel twirling, an optimal measurement for the ensemble $S_{0}^{(\mathcal{TN})}$ is given by $\{ M_{\x}^{\perp} \} _{\x=1}^n$ where $M_{\x}^{\perp}  = w_{\x} s (-\vec{m}_{\x}) $. \\

{\bf Proof.} For an ensemble $S_{0}^{(\i)}$, an optimal measurement can be found by maximizing the guessing probability
\bea
\max & & \frac{1}{n} \sum_{\x=1}^{n} \tr(M_\x \rho_\x) \nonumber 
\eea
with the constraint that $ M_\x\geq 0 \text{ and } \sum M_{\x} = {\bf I}$. Let $M_{\x} = w_{\x} s (\vec{m}_{\x}) $  denote POVM elements with $w_{\x} \in [0,1]$ and $s (\vec{m}_{\x}) = ({\bf I} + \vec{m}_{\x} \cdot \vec{\sigma})/2$. 

Then, the optimization problem is equivalently written as 
\bea
\max  & & \frac{1}{n} + \frac{1}{2n} \sum_{\x=1}^{n} w_\x \vec{m}_\x\cdot \vec{r}_\x  \label{eq:opt1}
\eea
where the constraints are $w_\x\geq 0$ and $ \sum_\x w_\x \vec{m}_\x\cdot \vec{\sigma} ={\bf I}$. Suppose that by twirling a channel $\N$ the fraction is given by $1-\eta_{\N}< 0$. The optimal discrimination for a resulting ensemble $S^{(\mathcal{TN} )}$ can be found by solving 
\bea
\max~~~ \frac{1}{n} + \frac{1-\eta_{\N}}{2n} \sum_{\x=1}^{n} w_\x \vec{m}_\x\cdot \vec{r}_\x. \label{eq:opt2}
\eea
Solutions of two optimization problems in the above are related by a simple inversion. The optimal measurement in Eq. (\ref{eq:opt2}) can be obtained by converting the direction $\vec{m}_\x \rightarrow -\vec{m}_\x$ for solutions $\vec{m}_\x $ in Eq. (\ref{eq:opt1}). $\Box$\\

In particular, for two-state discrimination, the update of a measurement is simply re-labelling of POVM elements since two POVM elements are orthogonal to each other. This shows an analogy to the aforementioned classical counterpart.

\subsection{ Twirling a channel in two-state discrimination}

For an ensemble of two states, the protocol of channel coding of a measurement in Sec. \ref{sec:chm} works not only for cases of equal {\it  a priori} probabilities but also when {\it  a priori} probabilities are unequal. \\

{\bf Proposition.} A measurement for an ensemble of two states can be generally preserved by an arbitrary channel.\\

{\bf Proof.} For an ensemble, $S^{(\i)} = \{q_{\x} ,\rho_{\x} \}_{\x=1}^2$, an optimal measurement can be found by finding the spectral decomposition,
\bea
&& q_1 \rho_1 - q_2 \rho_2 = r_2 \sigma_2 - r_1 \sigma_1 \nonumber 
\eea
where $\sigma_1$ and $\sigma_2$ are a pair of orthogonal pure states and $r_1$ and  $r_2 $ are non-negative. It also holds that $\sigma_1 + \sigma_2 ={\bf I}$.  Then, optimal POVMs are $M_1 =\sigma_{1}^{\perp}$ and $M_2 =\sigma_{2}^{\perp}$. 

Suppose that a channel $\N$ is twirled by the protocol in Sec. \ref{sec:chm} so that a depolarization channel in Eq. (\ref{eq:depn}) is obtained. To find an optimal measurement for the resulting ensemble, we find the spectral decomposition, 
\bea
&& q_1 \D_{ \eta_{\N}} [\rho_1] - q_2 \D_{ \eta_{\N}} [\rho_2] \nonumber \\
&& = (1-\eta_{\N}) (r_2 \sigma_2 - r_1 \sigma_1 ) + \eta_{\N} (r_2 - r_1)  {\bf I}/2.  \nonumber
\eea
From the relation $\sigma_1 + \sigma_2 ={\bf I}$, we have 
\bea
q_1 \D_{\N,\eta} [\rho_1] - q_2 \D_{\N,\eta} [\rho_2]  = \widetilde{r}_{2} \sigma_2 -  \widetilde{r}_{1} \sigma_1, \nonumber
\eea
where $\widetilde{r}_{i} = r_i - (r_1 + r_2) \eta_{\N} /2 $. Thus, it is shown that a measurement is preserved. $\Box$ \\
  
Note that for an ensemble $S^{(\i)} = \{q_{\x} ,\rho_{\x} \}_{\x=1}^2$, a channel $D_{\eta_{\N}}$ does not fulfil the OMP condition in Eq. (\ref{eq:popt}) although it is OMP as shown above. This confirms that the condition in Eq. (\ref{eq:popt}) is only sufficient for a channel to preserve an optimal measurement.

\subsection{Twirling a channel for multiple states} 

For more than two states, the protocol of channel coding of a measurement in Sec. \ref{sec:chm} fails to preserve a measurement if {\it  a priori} probabilities are unequal. We here provide an example that twirling a channel fails to preserve a measurement. 

Let us consider a set of modified trine states with unequal {\it a priori} probabilities $S^{(\i)}=\{q_\x ,\rho_\x\}_{\x=1}^{3}$, with $q_1=2q_2=2q_3= 1/2$. Trine states are three states that are equally spaced in a half-plane of a Bloch vector. Let $\vec{r}$ denote the Bloch vector of a qubit state $\rho$. Then, we consider modified trine states having Bloch vectors $\vec{r}_1 = ( 1/2,0,0)$, $\vec{r}_2 = ( -1/2, \sqrt{3} / {2},0)$ and $\vec{r}_3=(- {1}/{2},- {\sqrt{3}}/{2},0)$, i.e., the first state is not pure. For the ensemble, an optimal measurement can be found explicitly \cite{ha2013} as follows 
\begin{align}
&\left\{
\begin{pmatrix}
0.44 & 0.44 \\
0.44 & 0.44 \\
\end{pmatrix} \,,
\begin{pmatrix}
0.28 & -0.22-0.17 i \\
-0.22+0.17 i & 0.28 \\
\end{pmatrix}\,, \right. \notag \\
& \left. \qquad 
\begin{pmatrix}
0.28 & -0.22+0.17 i \\
-0.22-0.17 i & 0.28 \\
\end{pmatrix}
\right\}. \nonumber 
\end{align}
Note that none of POVM elements are non-zero.

Suppose that the ensemble is sent through a channel $\N$ with the protocol of channel coding in Sec. \ref{sec:chm}. We assume that a depolarization channel is obtained with $1-\eta_{\N}=2/3$. For the ensemble $S^{(\mathcal{TN})}$, an optimal measurement can be found as follows
\begin{align}
&\left\{
\begin{pmatrix}
0.46 & 0.46 \\
0.46 & 0.46 \\
\end{pmatrix} \,,
\begin{pmatrix}
0.27 & -0.23-0.14 i \\
-0.23+0.14 i & 0.27 \\
\end{pmatrix}\,, \right. \notag \\
& \left. \qquad 
\begin{pmatrix}
0.27 & -0.23+0.14 i \\
-0.23-0.14 i & 0.27 \\
\end{pmatrix}
\right\}. \nonumber 
\end{align}
Thus, it is shown that an optimal measurement is not preserved with the protocol of channel coding. Note that the {\it a priori} probabilities are not equal.

\section{Simulation on a quantum computer}

In this section, we present a proof-of-principle demonstration of channel coding of a qubit measurement with the IBM quantum computer ibmqx2 using the Qiskit package \cite{aleksandrowicz2019}. We specifically consider ensembles of a pair of orthogonal states $S_Z = \{\ket{0},\ket{1}\}$ and the four-states in the Bennett-Brassard 1984 (BB84) protocol \cite{bennett1984} $S_\textrm{BB84} = \{\ket{0},\ket{1},\ket{+},\ket{-}\}$, both with equal {\it a priori} probabilities. We consider flip channels
\bea
	\mathcal{N}_R [\rho] = (1-p_{f})\rho + p_{f} R \rho R \label{eq:chex}
\eea
where $p_{f} \in [0,1]$ is a flipping probability and $R = X, Y$ where $X$ and $Y$ are the Pauli operators, which arise as a result of the interactions of the system qubit $\rho$ with an environment.

\begin{figure}
	\centering
	\includegraphics[width = 0.48\textwidth]{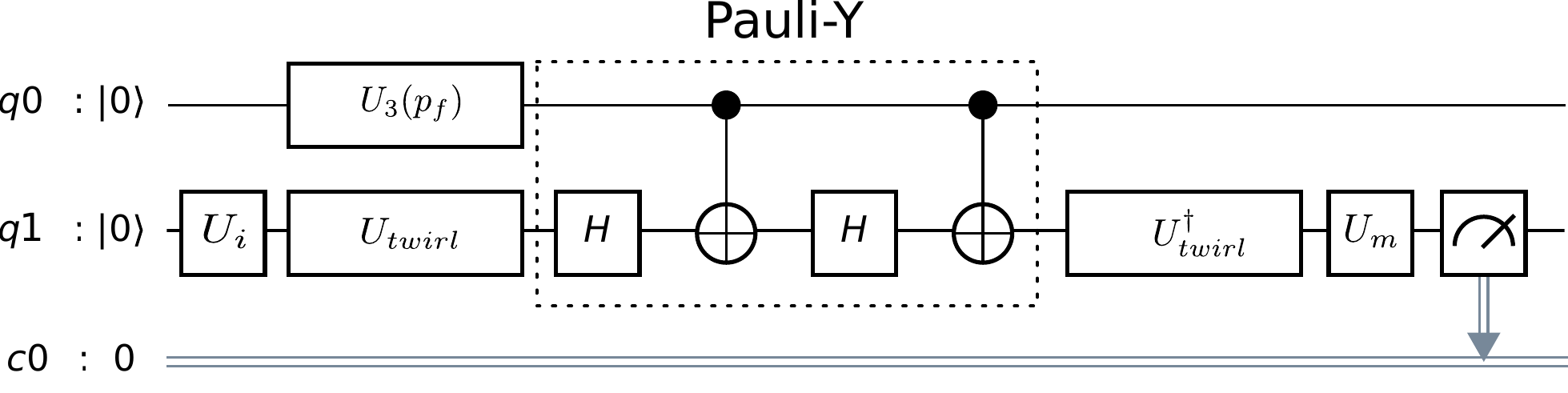}
	\caption{ A quantum circuit for channel coding of a qubit measurement is shown. The interaction between a system qubit q1 and an environment (aniclla) qubit q0 in Eq.~(\ref{eq:cY}) is realized with two CNOT gates, so that the channel for the system is described by $\N_Y$. By default all qubits are initialized in $|0\rangle$, so that a unitary operation $U_i$ is required to prepare other states, for instance $|1\rangle$, $|+\rangle$, and $|-\rangle$. Analogously, since measurements are always done in the $Z$-basis, a unitary transformation $U_m$ is required to prepare POVM elements other than $|0\rangle \langle 0|$ or $|1\rangle \langle 1|$. The channel $\N_X$ is realized by replacing the block inside the dashed box with a single CNOT gate. The gates $U_{twirl}$ denote the application of one of the elements in a unitary $2$-design, Eq.~\eqref{eq:u2}. 
	}\label{fig:quantumcircuits}
\end{figure}

The aforementioned channels can be realized with a quantum circuit with two qubits -- q1 for the system , and q0 the ancillary qubit mimicking an environment -- see Fig.~\ref{fig:quantumcircuits}. To this end, we need gate operations, Pauli $X$, Hadamard $H$, and controlled-not (CNOT) gate denoted by $C_X$. Note that with the Qiskit notation, the most general single-qubit operation is written as
\bea
U_3(\theta, \phi,\lambda)  = \begin{pmatrix}
	\cos(\theta/2) & -e^{i\lambda} \sin(\theta/2)\\
	e^{i\phi}\sin(\theta/2)  & e^{i\lambda +i\phi}\cos(\theta/2)
	\end{pmatrix}. \label{eq:siq}
\eea
In order to obtain the channel $\N_{X}$ on the system q1, the environment q0 is prepared in the state  
\begin{align}
	\ket{a}_E = U_3(2\alpha,0,0) \ket{0}_E = \cos(\alpha)\ket{0}_E + \sin (\alpha)\ket{1}_E, \nonumber
\end{align}
where the angle $\alpha$ corresponds to the flipping probability as $\cos(\alpha)^2 = 1-p_f $. Then, when the system is prepared in a state $|\psi\rangle$, the interaction with the environment is described as 
\bea
	C_{X} \ket{a}_E \ket{\psi} = \sqrt{1-p_f}\ket{0}_E \ket{\psi} + \sqrt{p_f}\ket{1}_E  X\ket{\psi}. \label{eq:Cnot}
\eea
The channel $\N_{X}$ is obtained by tracing out the environment qubit q0, which in practice is done by measuring q1 only. For an observable $M$ of interest (in particular POVMs) on the system, the expectation is given by
\bea
\tr[M \N_{X} [|\psi\rangle \langle \psi|]] = \tr[ ({\bf I} \otimes M ) ~C_{X } | a \rangle_E \langle a| \otimes | \psi\rangle \langle \psi| C_{X}^{\dagger} ] \nonumber
\eea
Similarly, the channel $\N_Y$ is obtained by implementing the two-qubit controlled-Y gate $C_Y$ as follows (see also Fig.~\ref{fig:quantumcircuits})
\bea
C_{Y}  \ket{a}_E  \ket{\psi}  =	\sqrt{1-p_f}\ket{0}_E  \ket{\psi} + \sqrt{p_f}\ket{1}_E  Y\ket{\psi}, \label{eq:cY}
\eea
where $ C_{ Y }  = C_{ X } ({\bf I} \otimes H) C_{ X} ({\bf I} \otimes H)$ and subsequent tracing out the environment qubit. 

In order to implement channel coding of qubit measurement, the minimal unitary 2-design $W$ in Eq.~\eqref{eq:u2} is realized with the unitary gate $U_2(\phi, \lambda) =U_3(\pi/2,\phi,\lambda)$ as
\bea
 && \bigl\{U_2(0,\pi/2),U_2(\pi,\pi/2),U_2(0,3\pi/2),U_2(\pi,2\pi/2),  \nonumber \\ 
&& U_2(\pi/2,\pi),U_2(\pi/2,0),U_2(3\pi/2,\pi),U_2(3\pi/2,0) \bigr\}, \nonumber
\eea
together with ${\bf I}$, $-iX$, $-iY$, and $-iZ$. Note that for the ensemble $S_Z$, a measurement in the basis $Z$ is optimal. With this measurement, for a Pauli channel $\mathcal{N}_X$ we have (see also Eq. (\ref{eq:guessp}))
\bea
	p_\textrm{guess}^{(\mathcal{N}_X)} = \frac{1}{2} + \frac{|1-2p_{f}|}{2},~\mathrm{and}~p_\textrm{guess}^{(\mathcal{T} \mathcal{N}_X)} = \frac{1}{2} + \frac{|3-4p_{f}|}{6}. ~~~~~\label{eq:theo2N}
\eea
For the ensemble $S_\textrm{BB84}$ a random measurement in the $X$ and $Z$ basis are optimal. With the measurement, we have 
\bea
	p_\textrm{guess}^{(\mathcal{N}_Y)} = \frac{1}{4} + \frac{|1-2p_{f}|}{4},  ~\mathrm{and}~p_\textrm{guess}^{(\mathcal{T} \mathcal{N}_Y)} =  \frac{1}{4} + \frac{|3-4p_{f}|}{12}. ~~~~~\label{eq:theo4N}
\eea

To implement the twirling protocol, we collect the data for 8000 (maximum allowed is 8192) shots for each unitary matrix in the unitary-2-design $W$, which is applied before and its conjugate after the channel $\mathcal{N}_R$, and perform the averaging in post-processing. 

As a proof-of-principle, this is equivalent to random applications of unitary $2$-design before and after a channel use. Thus, channel coding of a measurement is implemented. We show in Fig.~\ref{fig:results} the guessing probabilities Eq.~\eqref{eq:guessp} measured on the $5$-qubit quantum machine ibmqx2. We verified that the results from the classical quantum circuit simulator 'qsam\_simulator' are in perfect agreement with Eqs.~\eqref{eq:theo2N} and \eqref{eq:theo4N}. This certifies that the circuits are simulating correctly the quantum channel and the twirling protocol. 

\subsubsection*{Comments on the simulation} 
\begin{figure}
	\centering
	\includegraphics[width = 0.53\textwidth]{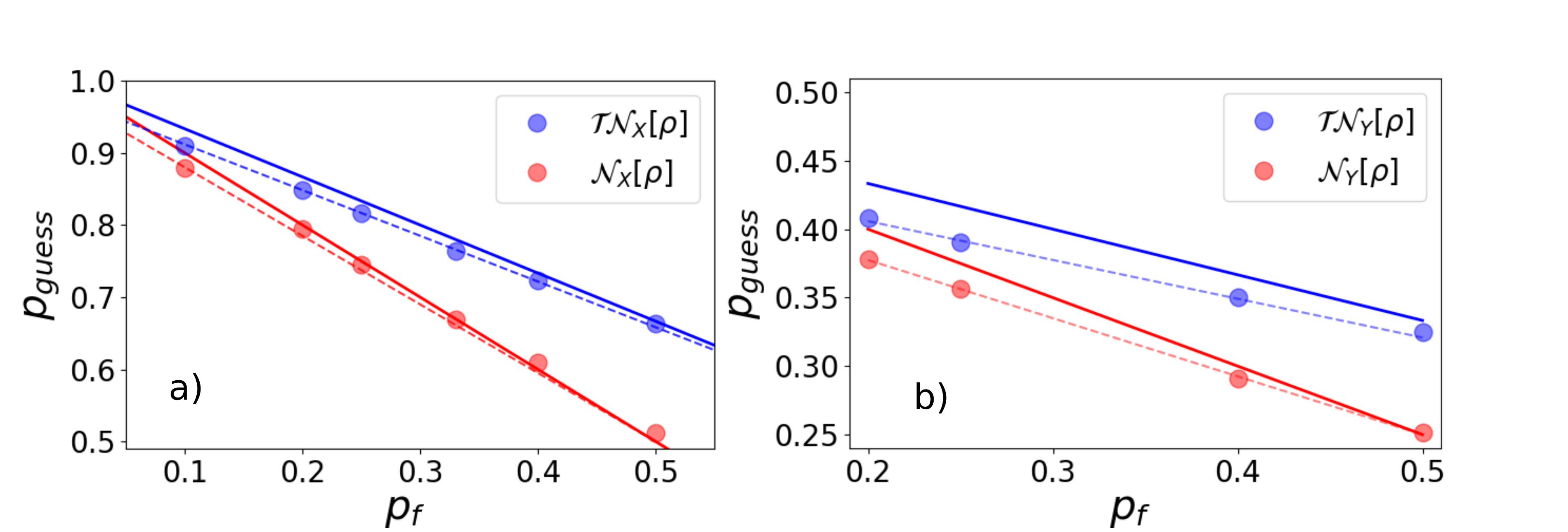}
	\caption{Guessing probabilities for channels $\N_X$ and $\N_Y$ in Eq. (\ref{eq:chex}) are shown: a) a channel $\N_X$ for a pair of orthogonal states $S_{Z}$ and b) a channel $\N_Y$ for the BB84 states $S_{\mathrm{BB84}}$. The solid lines are theoretical prediction in Eqs. (\ref{eq:theo2N}) and (\ref{eq:theo4N}). The guessing probabilities after a channel are in the red color. After twirling a channel, they are in the blue color. Circles are guessing probabilities from the ibmqx2. The circle size is larger than the statistical measurement noise. The dashed line is made by assuming the noise model in Eq.~\eqref{eq:MeasNoise} $\eta=0.05$ for a) and $\eta=0.15$ for b), which shows a reasonably good explanation about the source of errors in IBM quantum computers. }\label{fig:results}
\end{figure}

The results in Fig.~\ref{fig:results} obtained in March in 2019 show a good agreement with the theoretical prediction in Eqs.~\eqref{eq:theo2N} and \eqref{eq:theo4N}. A certain loss of probability can be observed that can be modelled by shot noise on the measurement outcome, 
\begin{align}\label{eq:MeasNoise}
	\tr(M \rho) \rightarrow \tr(M[(1- \eta)\rho+ \eta {\bf I}/2]), 
\end{align}
as indicated by the dashed line. The same measurements were performed in June 2019. It turns out that the shot noise was significantly higher ($\eta = 0.65$ instead of $\eta = 0.05$ for $S_Z$, $\eta =0.57$ instead of $\eta =0.15$ for $S_\textrm{BB84}$). This seems consistent with a recent study of the noise sources on the IBM quantum computers \cite{chen2019} where a loss of the overall norm of the Bloch vectors due to shot noise is identified as one of the major sources of error. The same measurements were performed in June on ibmqx4 and ibmq\_16\_melbourne machines. For the ibmq\_16\_melbourne, the shot noise was $\eta \approx 1$ so that no significant results could be extracted. For the ibmqx4, $p_\textrm{guess}$ for $\mathcal{T}\mathcal{N}_R$ was systematically larger than for $\mathcal{N}_R$, in agreement with the theory. However, the measured $p_\textrm{guess}$ values strongly fluctuated around the theoretical values. 
These observations cannot be accounted for by the estimates of the gate fidelities reported by IBM Q Experience.


In an attempt to find the sources of the errors, we measured the state of the environment qubit q0 to control the actual value of $p_f$. Some fluctuations were observed, which are correlated with the fluctuations in $p_\textrm{guess}$ on ibmqx4, but can only partially explain the deviations from the theory. Consequently, further understanding of the device's imperfections is of paramount importance for channel coding, and for quantum information processing in general, on available quantum hardware.

\section{Conclusion}

We have formulated and presented a framework of preserving a measurement for quantum systems interacting with an environment. We show that an LOCC protocol with local unitaries only can realize the preservation of an optimal measurement without further resources such as a larger Hilbert space, contrasting to the case of preserving states. A general framework for channel coding of a measurement is presented as a supermap that transforms a channel to an OMP one. In particular, it is shown that channel twirling implements the preservation of a measurement for ensembles of equally probable states. For qubit ensembles, the protocol of preserving a measurement is investigated in detail and is found that the protocol works for i) any pair of qubit states and ii) ensembles of equally probable states. A counter-example for three states is explicitly provided. Channel coding of a qubit measurement is demonstrated for ensembles of a pair of orthogonal states and the four states in the BB84 protocol, and can be readily applied to practical quantum communication protocols. 

Our work sheds a new light in directions of an early-stage quantum information processor, that works with limited resources and restricted controls. For instance, the preservation of quantum states in a noisy environment, which needs ancillas and a high-precision control over a system and ancillas, may not be achieved within a near future. The presented framework of preserving an optimal measurement could be a next best and feasible opportunity. Our demonstration with IBM quantum computers has shown that by channel coding of a measurement, single-qubit information processing can readily work against an adversarial environment that may cause high-rate errors. Our work initiates a new direction toward an early-stage quantum information processing, leading to a number of questions. First, channel coding of a measurement for ensembles of arbitrary {\it a priori} probabilities is sought, to apply channel coding of a measurement for arbitrary ensembles. Next, for further extension beyond a single qubit state, it is significant to find a minimal unitary $2$-design or its approximation for multiple qubits. In addition, it is interesting to find how tightly schemes of preserving states and a measurement are related to each other. In future investigations, it would be also interesting to apply channel coding of a measurement in a realistic and practical application such as few-qubit quantum algorithms.

\section*{Acknowledgment}
J.B. thanks V. Scarani, H.-K. Ng, and A. Winter for helpful discussions. This work is supported by National Research Foundation of Korea (NRF-2017R1E1A1A03069961), the KIST Institutional Program (2E29580-19-148) and ITRC Program(IITP2018-2019-0-01402), NASA Academic Mission Services (contract number NNA16BD14C), and Iniziativa Specifica 341 INFN-DynSysMath. The authors acknowledge the use of IBM QISKIT for this work. The views expressed are those of the authors and do not reflect the official policy or position of IBM. The authors acknowledge support from the NASA Advanced Exploration Systems program and the NASA Ames Research Center. The views and conclusions contained herein are those of the authors and should not be interpreted as necessarily representing the official policies or endorsements, either expressed or implied, of the U.S. Government. The U.S. Government is authorized to reproduce and distribute reprints for Governmental purpose notwithstanding any copyright annotation thereon.

\end{document}